\documentclass[fleqn,usenatbib]{mnras}

\usepackage[T1]{fontenc}
\usepackage{multirow}
\PassOptionsToPackage{unicode}{hyperref}
\PassOptionsToPackage{naturalnames}{hyperref}
\DeclareRobustCommand{\VAN}[3]{#2}
\let\VANthebibliography\thebibliography
\def\thebibliography{\DeclareRobustCommand{\VAN}[3]{##3}\VANthebibliography}

\usepackage{graphicx}	
\usepackage{amsmath}	
\usepackage{amssymb}
\usepackage{booktabs}
\usepackage{makecell}
\usepackage{multirow}
\usepackage{hyperref}

\title[VHE excess in nearby BL Lacs]{Understanding the Very High Energy $\gamma$-ray excess in nearby blazars  using  leptonic model}

\author[Aaqib Manzoor et al.]{
Aaqib Manzoor$^{1}$\thanks{E-mail: aqibmanzoor1111@gmail.com},
Sunder Sahayanathan$^{2,4}$\thanks{E-mail: sunder@barc.gov.in}, 
Zahir Shah$^{3}$\thanks{E-mail: shahzahir4@gmail.com},
Subir Bhattacharyya$^{2,4}$, 
Naseer Iqbal$^{1}$ \&
\newauthor
Zahoor Malik$^{1}$ \\
$^{1}$Department of Physics, University of Kashmir, Srinagar 190006, India.\\
$^{2}$Astrophysical Sciences Division, Bhabha Atomic Research Center, Mumbai 400085, India.\\
$^{3}$Department of Physics, Central University of Kashmir, Ganderbal 191201, India.\\
$^{4}$Homi Bhabha National Institute, Mumbai 400094, India.
}

\date{Accepted XXX. Received YYY; in original form ZZZ}

\pubyear{2022}

\begin{document}
\label{firstpage}
\pagerange{\pageref{firstpage}--\pageref{lastpage}}
\maketitle

\begin{abstract}
The availability of simultaneous X-ray and Very High Energy (VHE) observations of blazars helps to identify the plausible radiative contributors to the VHE emission. Under leptonic scenario, the VHE emission from BL Lacs are attributed to the synchrotron self Compton (SSC) emission. However, many BL Lacerate (BL\,Lacs) have shown significant hardening at VHE after correction for the Extra Galactic Background Light (EBL) attenuation. We study the spectral hardening of two nearby BL Lac objects, Mkn\,421 and
Mkn\,501 having most number of simultaneous X-ray and VHE observations available among all the blazars. These BL Lacs are relatively close and the effect of  EBL attenuation is relatively minimal/negligible. We study the scatter plot between the X-ray spectral indices
and intrinsic VHE indices to identify the plausible origin of the VHE emission. For Mkn\,501, the VHE spectral indices are 
steeper than X-ray spectra,  suggesting the scattering process happening at extreme Klein-Nishina regime. On the other hand, for Mkn\,421, the VHE spectra is remarkably harder than the X-ray spectra, which suggests an additional emission mechanism other than the SSC process.  We show this hard VHE spectrum of Mkn\,421 can be explained by considering the inverse Compton (IC) emission from a broken power law electron distribution with Maxwellian pileup.  The possibility of the hadronic contribution at VHE $\gamma$-rays is also explored by modeling the hard spectrum under photomeson process.
\end{abstract}

\begin{keywords}
galaxies:active -- quasars: individual: Mkn\,421,Mkn\,501 -- galaxies: jets -- radiation mechanism: non-thermal -- gamma-rays: galaxies 
\end{keywords}



\section{Introduction}

Blazars are subclass of active galactic nuclei (AGN) with powerful relativistic jet pointing close to the line of sight of observer \citep{Urry1995, a2010ApJ...715..429A, acero2015ApJS..218...23A}.  The jet emission is non-thermal in nature and dominates the overall spectrum of 
blazars. Further, the alignment of this relativistic jet results in significant enhancement of the flux due to Doppler effect, and
extreme observational features like rapid variability down to time-scales of minutes and super-luminal motion in high 
resolution radio maps \citep{Urry1995}. 
Blazars are broadly classified as flat-spectrum radio quasars (FSRQs) and BL\,Lacs 
based on the presence or absence of spectral line features in their optical spectra \citep{stickle1991ApJ...374..431S,Urry1995}.  

The broadband spectral energy distribution (SED) of blazars are characterized by a typical double hump feature with 
the low energy component peaking at optical/UV/X-ray band, while the high energy component at GeV/TeV energy. The low energy 
spectral component is well understood to be synchrotron emission arising from a non-thermal distribution of charged particles 
gyrating in  the magnetic field of the  jet. 
Based on the peak frequency $\nu_{sy}$ of the synchrotron component,  BL Lac objects are  further classified into high energy peaked 
blazars (HBL) ($\nu_{sy} \ge 10^{15}$ Hz), intermediate peaked blazars (IBL) ($10^{14} \leq \nu_{sy} \leq 10^{15}$ Hz) and low 
energy peaked blazars (LBL) ($\nu_{sy} \le 10^{14}$ Hz) \citep{f1998MNRAS.299..433F}.

The origin of the high energy component is still unclear with the inverse Compton 
up-scattering of low energy photons by a relativistic electron distribution being the most 
competitive model (leptonic model) \citep{1985A&A...146..204G, 1993ApJ...416..458D, 1994ApJ...421..153S, 2002A&A...384...56C,   2018MNRAS.477.4749C}. If the synchrotron photons 
itself gets up-scattered to $\gamma$-ray energies then the scattering process is termed as 
synchrotron self Compton (SSC) emission (\citet{Ghisellini_1998}). On the other hand, if the soft photons are of external origin 
then the emission process is termed as external Compton (EC).  The plausible external photon field participating in the EC process
can be the IR photons from the dusty torus \citep{dustyApJ...545..107B}, broad emission lines from the hot gases (broad line region) 
\citep{blr1993ApJ...416..458D} and the thermal photons from the accretion disk \citep{1993ApJ...416..458D, 1997A&A...324..395B}.  In case of HBLs, the $\gamma$-ray emission  can be understood
within the framework of SSC emission models \citep{2001A&A...367..809K, 2011ApJ...736..131A, 2015JHEAp...5...30A, 2021MNRAS.505.2712D}, while the $\gamma$-ray emission from IBL/LBL/FSRQ support the EC process \citep{2012ApJ...754..114H, 2013ApJ...768...54B, 2015A&A...578A..69H}. High energy component of  the  SED can also 
be explained by the hadronic emission models considering proton synchrotron process and/or pion decay 
processes \citep{1993A&A...269...67M,Aharonian_2000, cerruti/mnras/stu2691}. The leptonic models are mostly  favoured
over hadronic models due to large jet power required by the later \citep{bou2013EPJWC..6105003B, der2016ApJ...825L..11P}. 
Nevertheless, detection of neutrino from AGN and orphan VHE flares strongly favours the models involving hadronic
interaction \citep{2022ApJ...927..197A}.

Blazars comprise majority of the extragalactic $\gamma$-ray sources in the universe \citep{wak2008ICRC....3.1341W}. 
The VHE $\gamma$-ray emission from blazars en-route to the observer gets absorbed by  EBL via pair production process, 
 which causes the observed VHE spectrum to differ from the intrinsic VHE spectrum  \citep{Salamon_1998}. For the sources at larger redshift, the VHE flux gets significantly attenuated by EBL and the resultant observed spectrum will be steeper than the intrinsic one.  In order to obtain the  physical parameters responsible for the VHE emission,  one needs to obtain the underlying intrinsic spectrum. 
 In most of VHE blazars, the EBL correction to the observed spectrum results in the harder intrinsic spectrum, which poses a challenge for the conventional emission models; nevertheless, attempts have been made to explain the hard 
VHE spectra.
The hardening can be attributed to the internal absorption of $\gamma$-rays due to interactions with dense and narrow radiation fields located near the compact emission regions \citep{Aharonian_2008}. \citet{0.710.1111/j.1745-3933.2006.00156.x} has shown that a harder $\gamma$-ray spectrum can be produced under SSC emission scenario when the underlying electron distribution is very hard with high minimum energy cut-off. 
A hard electron distribution with the index less than 2 (corresponding inverse Compton spectral index less than 1.5) can be attained if the electrons are accelerated at relativistic shock fronts \citep{relation2007}. 
Addition to this, during one of the major VHE flare of Mkn\,501, the source also showed an unusual excess at 3 TeV which can be fitted by a combination of log-parabola and and exponential function \citep{2020A&A...637A..86M}. Such a spectrum can be reproduced by considering an electron distribution which is a combination of a power-law function with a pile up \citep{wenhu2021narrow}.

VHE hardening in blazars and the underlying physics can be best understood by studying the nearby bright  blazars, as the effect of EBL on their spectrum will be minimum.  Mkn\;421 and Mkn\;501 are the two brightest nearby BL Lac objects and are among  few blazars with  good quality data available for complete broadband SED modeling \citep{abdo2011ApJ...736..131A}. Mkn\;421 was also the first blazar discovered at TeV energy by Whipple telescope in 1992 \citep{1992Natur.358..477P, fossotiApJ...677..906F} and one of the closest BL Lac objects located at a redshift of 0.0308  \citep{ulrich1975ApJ...198..261U}.  Mkn\,501 at redshift 0.034 is another intensively monitored nearby BL Lac object  \citep{abdo2011ApJ...736..131A,2012-501Ahnen_2018} and the  VHE emission from this source was first observed  by the Whipple observatory in 1995 \citep{Quinn__1996}. Among blazars, Mkn\,421 and Mkn\,501 also have maximum number
 of simultaneous X-ray and VHE observations available. Significant number of observations and being nearby blazars makes them the ideal laboratory for studying the VHE spectral properties. In this paper, we examine the VHE spectral hardening of Mkn\,421 and Mkn\,501 in comparison with their simultaneous X-ray spectrum using archival \emph{MAGIC} and \emph{Swift}-XRT observations. Further, we supplement the study with the broadband spectral modelling of the hard VHE spectrum considering different emission scenario.
The paper is organised as follows:  In section  \S \ref{sec:2},  we give the details of the observations and the data analysis procedure and in \S \ref{sec:3} the hard VHE spectra is studied. In \S \ref{sec:4}, we perform the broadband spectral fitting of the source with relatively hard VHE spectrum and summarise the results in \S \ref{sec:5}. Throughout the work we use a cosmology with $\rm \Omega_M = 0.3$, $\rm \Omega_\Lambda = 0.7$, and $\rm H_0 = 71  km s^{-1} Mpc^{-1}$.

\section{Observations and Data Reduction}
\label{sec:2}

The simultaneous broadband information of the Bl Lac objects Mkn\,421 and Mkn\,501 in  Optical/UV, X-ray, $\gamma$-ray and VHE energy bands are obtained from 
\emph{Swift}-UVOT/XRT, \emph{Fermi}-LAT, and  Major Atmospheric Gamma-ray Imaging Cherenkov Telescope (\emph{MAGIC}) observations. For the VHE hardness 
study, we acquire the archival spectral information in the X-ray and VHE band from the 
literature (\cite{2008refId0}; \cite{2013Balokovi__2016}; \cite{2012-501Ahnen_2018};\cite{2014-5012020}) whereas, for the broadband spectral
fitting we reduce the Optical/UV and X-ray data from \emph{Swift} observations, and the $\gamma$-ray data from \emph{Fermi} observations.

The blazar Mkn\,421 was observed intensively during December, 2007 -- June,  2008 by \emph{MAGIC} telescope $(E > 100\,GeV)$  and  \emph{Swift}
observatory in the optical/UV and X-ray bands.  
During this period, the source was in active state, and 
eight VHE observations have simultaneous data available in the X-ray band \citep{2008refId0}. The VHE and X-ray spectral shape during these eight 
epochs were represented either by a log-parabola or a power-law functions. Unlike power-law, the spectral slope in case of log-parabola will be energy dependent
and hence cannot be used readily for the hardness study. Therefore, we use only those simultaneous observations for which the X-ray and the VHE spectra 
can be well represented by a power-law function. Under this criteria, only three out of eight observations were selected during this observation window.
The next multi-wavelength observation of Mkn\,421 was performed in 2013 during which the source was in quiescent state. 
The detailed analysis of the data during this observation window was reported in \cite{2013Balokovi__2016} and with eleven simultaneous X-ray and VHE 
spectral fits. For nine epochs, the X-ray and VHE spectra were well represented by a power-law function and these observations are 
considered for the present study. The observed VHE spectrum is corrected for EBL induced absorption using the opacity estimates by \citet{Franceschini_2008} 
and \citet{Dom_nguez_2010}. Since at low redshifts these EBL models are consistent, the predicted intrinsic source VHE spectra do not vary significantly (see Figure~\ref{Fra_Dom}).
In Table~\ref{tab:table}, we provide the details of the simultaneous X-ray and VHE observations of Mkn\,421.

Mkn\,501 is extremely high energy peaked blazar (EHBL) and has been the target of simultaneous observations during various flux states \citep{2012-501Ahnen_2018}.
The source was observed in 2012,  and  during this period eight simultaneous X-ray and VHE spectra were available \citep{2012-501Ahnen_2018}. All 
the spectra can be well represented by a power-law function and are included for the present study.
The source was again observed simultaneously at different energy bands for nearly two weeks of July, 2014. 
During this period, the enhanced X-ray flux let to have fourteen simultaneous X-ray and VHE observations \citep{2014-5012020}. 
However, only nine spectra followed a power-law function and are used for this study. The details of the selected epochs are given in Table~\ref{tab:table}.
We have not performed EBL correction for these VHE observations instead use the EBL corrected VHE spectra reported in the 
original works \citep{2012-501Ahnen_2018,2014-5012020}. 

Among these two sources, Mkn\,421 shows hard VHE spectra during different flux 
states \citep{2008refId0, 2013Balokovi__2016}.
During the 2013 multi-wavelength campaign of Mkn\,421, the maximum hardness in VHE spectrum was observed on 
MJD 56335 \citep{2013Balokovi__2016}. We construct simultaneous broadband SED of the source during this epoch to 
examine whether this hardness is consistent with the leptonic interpretation of the VHE spectrum (\S \ref{sec:4}). 
The broadband SED is obtained by supplementing the VHE observation with the \emph{Swift}-XRT/UVOT and 
\emph{Fermi}-LAT $\gamma$-ray observations.

\subsection{Data Reduction}
\label{sec:2.1}

\subsubsection{Fermi-LAT}
 The \emph{Fermi} Large Area Telescope (\emph{Fermi}-LAT) is one of the two instruments onboard the \emph{Fermi} Gamma Ray Space
Telescope launched in 2008. %
\emph{Fermi}-LAT  operating in all-sky scanning mode observes the galactic and extra galactic sources in the energy range 20 MeV to 500 GeV \citep{atwood2009ApJ...697.1071A}.
Since 2008, the \emph{Fermi}-LAT has been  monitoring blazars continuously with a broad field of view (FoV) $\sim 2.4$ sr \citep{atwood2009ApJ...697.1071A}. In this work,  we select source class events in the energy range 0.1--300 GeV  within 15 degree  of the position of the source.  We have used Pass 8 data with the photon-like events  classified as
’evclass=128, evtype=3’. The data is converted to science products using the \emph{Fermitools} (formally Science Tools) with the latest version  v2.0.1 \footnote{\url{https://fermi.gsfc.nasa.gov/ssc/data/analysis/documentation/}}. 
The recommended criteria
$``(DATA_-QUAL>0)\&\&(LAT_-CONFIG==1)"$ was used for the good time interval selection. The contamination from the $\gamma$-rays of bright
Earth limb  is avoided by using  a zenith angle cut of 90 deg. The Galactic diffuse and the isotropic emission components were modelled with the post-launch
instrument response function P8R3-SOURCE-V2 and
iso-P8R3-SOURCE-V2-v1.txt, respectively. In the
XML model file, we included all the sources from the 4FGL
catalogue which were within (15+10) deg ROI of the source location.
In the generation of $\gamma$-ray spectrum, we considered the source to be detected if the test statistics (TS) is > 9 which corresponds to a 3-sigma detection (\citet{mat1996MmSAI..67..607M}).

\subsubsection{Swift-XRT}
\label{Sec:2.1.2}
\emph{Swift} is a multi-wavelength telescope, specially designed to observe the transitory phenomena in the galactic and extra galactic sky. It is equipped with three instruments namely X-ray telescope (XRT), an ultraviolet and optical telescope (UVOT) and burst alert telescope (BAT).  During the time MJD 56335, Swift-XRT observed the source in the Window Timing (WT) mode.
We processed the WT data by following the standard data analysis procedure mentioned in the \emph{Swift} analysis thread page \footnote{https://www.swift.ac.uk/analysis/xrt/}.  The clean event file was produced by running XRTPIPELINE using the latest version of CALDB. In order to obtain the source spectra,  we used the XSELECT tool  available with  HEASOFT-6.24. The source spectra is extracted from a circular region of radius $\sim 20$ pixels (47.2 arcsec) centered at  the position of the source, while the background spectra is extracted from a circular region with the  radius 100 arcsec but from a neighbouring region excluding the source pixels. We used the standard grade selection 0--12. XRTMKARF tool  was used to produce the auxiliary response files (ARF). The 0.3--10 keV source spectrum was binned using the grppha tool  prior to spectral fitting to ensure that each bin had at least 20 counts. The spectrum was fitted in XSPEC by power-law model and the absorption along the light path was adjusted by using the Galactic value for neutral hydrogen column density as $N_{H}= 1.92 \; \times \;10^{20} cm^{-2}$ \citep{kerbala2005A&A...440..775K}.

\subsubsection{Swift-UVOT}
UVOT is a diffraction limited 30 cm optical--UV telescope co-aligned with the X-ray telescope. It is equipped with six distinct filters sensitive in the $1700-6500$ nanometer(nm) wavelength range with a field of view  $\sim 17^\prime\times 17^\prime$. The \emph{Swift}-UVOT observations are available  in UM2, UW1 and UW2 filters during the time MJD 56335. We checked these observations for aspect correction using ASPCORR, and subsequently made the correction by running {\it uvotskycorr}. Magnitudes were extracted from images using tool UVOTSOURCE with source and background regions of radii 5 and 10 arcsec respectively. We used the $E(B - V) = 0.0174$ mag and the ratio $A_{V}/E(B - V) = 3.1$ to correct the observed magnitudes for Galactic extinction.

\section{X-ray--VHE Spectral Hardeness} 
\label{sec:3}

The correlated variability observed between the X-ray and the $\gamma$-ray energy bands of Mkn\,421 and Mkn\,501 suggests the same electron population is responsible for the emission at these energies. For these sources, the X-ray emission is attributed to the synchrotron emission process and the VHE emission to SSC process. If the Compton scattering process happens in the Thomson regime then the spectral indices at these energies will be equal; while, under Klein-Nishina regime then the VHE spectrum will be steeper than the X-ray spectrum. However, if the VHE index is harder than the X-ray spectrum then it indicates additional emission process may be contributing at VHE.

To examine this, we compare the X-ray and VHE intrinsic spectral indices of Mkn\,421 and Mkn\,501 during its simultaneous observation periods. The spectral information in the X-ray and VHE band has been acquired from the literature (See  section  \S \ref{sec:2}). Blazar spectra in the VHE band are attenuated en route through pair production process with the EBL and hence the observed VHE spectrum should
be corrected to obtain the intrinsic spectrum \citep{Franceschini_2008}. 
For Mkn\,501, we consider the intrinsic VHE spectra reported in the original literature \citep{2012-501Ahnen_2018, 2014-5012020} while for Mkn\;421, the observed VHE spectrum reported in the literature \citep{2008refId0, 2013Balokovi__2016} are corrected for the EBL induced absorption considering the source spectrum to be a power law,
\begin{align}
\Gamma_{\rm int} = \Gamma_{\rm obs}- \frac{d\tau(E_*,z)}{d(\ln{E})}
\end{align}
where, $\Gamma_{\rm int}$ and $\Gamma_{\rm obs}$ are the intrinsic and observed 
VHE spectral indices, and $\tau(E_*,z)$ is the EBL opacity for the $\gamma$-ray photon of energy $E_*$ emitted by the source located at   
redshift $z$. We estimate $\tau(E_*,z)$ considering two different EBL models, \citet{Franceschini_2008} and \citet{Dom_nguez_2010}, and the differentiation 
is performed numerically.
For the differentiation, we linearly interpolated the opacity estimates provided as two dimensional array in \citet{Franceschini_2008} and \citet{Dom_nguez_2010}.
The $\Gamma_{\rm int}$ is evaluated at $E_*\sim 1\,TeV$ which is then compared with the X-ray spectral index. In Table~\ref{tab:table}, we provide the X-ray 
and the intrinsic VHE spectral indices for Mkn\,421 and Mkn\,501.
We find that the choice of EBL models used do not vary the estimated
$\Gamma_{\rm VHE}$ significantly and this is consistent with our earlier study \citet{z2022MNRAS.511..994M} where different EBL 
models agree to each other at low redshifts. 

\begin{figure} 
    \includegraphics[width=0.5 \textwidth]{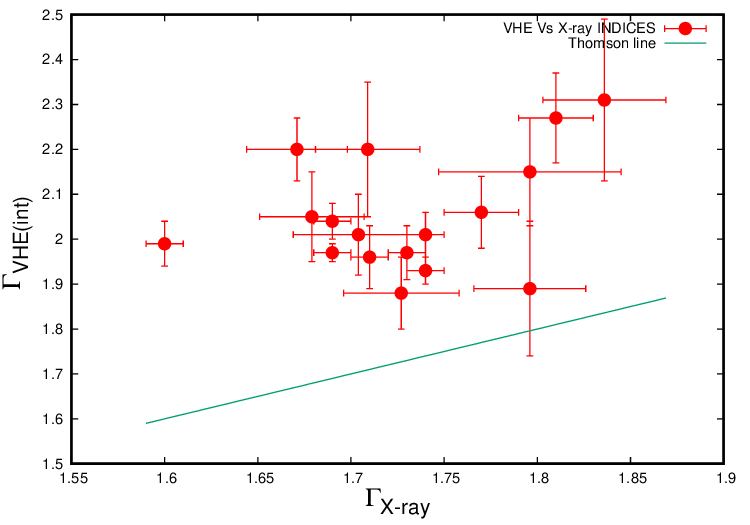}
 \caption{Scatter plot between the simultaneous X-ray (\emph{Swift}-XRT) spectral indices  and VHE (\emph{MAGIC}) intrinsic indices for Mkn\,501}
    \label{fig:fig1}
\end{figure}

\begin{figure} 
\includegraphics[width=0.5\textwidth]{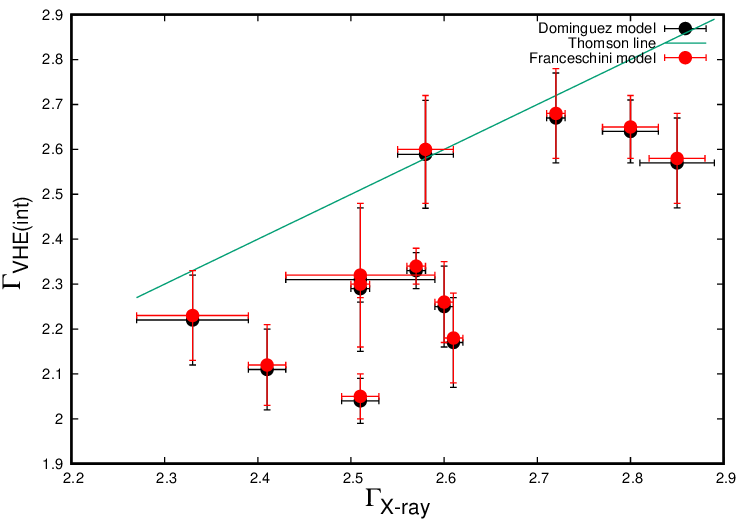}
\caption{Scatter plot between the  simultaneous X-ray (\emph{Swift}-XRT) spectral indices and VHE (\emph{MAGIC}) intrinsic indices for Mkn\,421 using \citet{Franceschini_2008} and \citet{Dom_nguez_2010}.}
\label{Fra_Dom}
\end{figure}

\begin{figure}
    \includegraphics[width=0.5\textwidth]{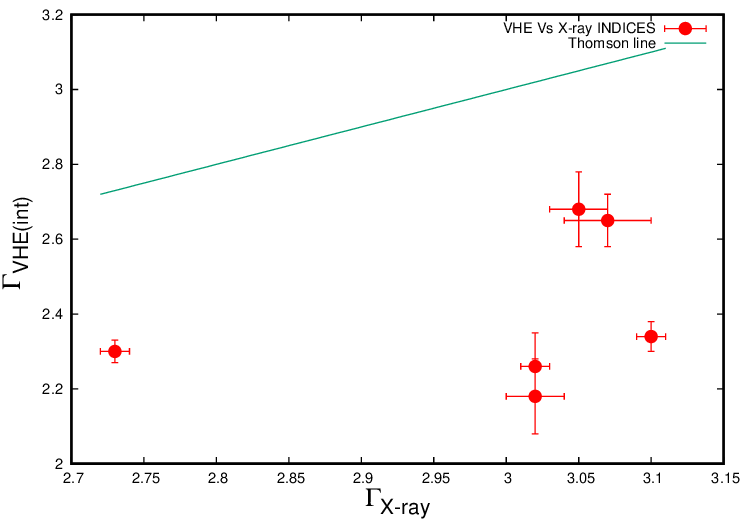}
        \caption{Scatter plot between the simultaneous X-ray \emph{(NuSTAR}) spectral indices  and VHE (\emph{MAGIC}) intrinsic indices for Mkn\,421}
            \label{fig:fig3}
\end{figure}

The scatter plot between the simultaneous X-ray spectral indices ($\Gamma_X$) and the intrinsic VHE spectral 
indices ($\Gamma_{\rm VHE}$) for the case of Mkn\,501, obtained from \emph{Swift}-XRT and \emph{MAGIC} observations
is shown in Figure~\ref{fig:fig1} along with the identity line. All the points fall
above the identity line indicating the VHE spectral indices are steeper than the corresponding simultaneous
X-ray spectral indices. If same electron population is responsible for the emission at these energies, then
this result supports the SSC scattering process to happen at Klein-Nishina regime. In Figure~\ref{Fra_Dom}, we show the 
similar plot obtained for the case of Mkn\,421 along with the identity line. Interestingly, all the points
fall below the identity line indicating a hard VHE spectra. We performed the study for the two EBL models 
mentioned earlier (black and red points) and the results are consistent with hard VHE spectra. This result cannot
be perceived simply under SSC scenario if same electron distribution is responsible for the emission at these
energies. We repeated this study by considering the hard X-ray spectral index obtained from \emph{Nu}-STAR
observations of the source  {\citep{2013Balokovi__2016} and the $\Gamma_{\rm VHE}$ obtained from \emph{MAGIC} observations (Figure~\ref{fig:fig3}).
Again, the data points show severe deviation from the identity line indicating a hard VHE spectra. Therefore, this study suggests that the VHE spectra of Mkn\,421 may also involve different emission component probably in addition to the SSC emission.

}
\begin{table*}
  \centering
\caption{Simultaneous X-ray and VHE observations of Mkn\,421 and Mkn\,501. Col 1: source name,  Col 2: time of X-ray observation, Col 3: power-law X-ray spectral indices, Col 4: time of VHE observation, Col 5: power-law VHE spectral indices, and Col 6: References.}
\begin{tabular}{|c|c|c|c|c|c|}
\hline
\multirow{2}{*}{Source Name} & 
\multicolumn{2}{c|}{X-Ray Observations} &
\multicolumn{2}{c|}{VHE observations} &
\multirow{2}{*}{Refrences}\\

\cline{2-5}
 & Time (MJD) & $\Gamma_x$ & Time (MJD) & $\Gamma_{VHE}$ \\
 
\hline
Mkn\,421  & 56302.1557 & 2.80 $\pm$ 0.03 & 56302.1365 & 2.65 $\pm$ 0.07 & \cite{2013Balokovi__2016}   \\
--  & 56307.3519 & 2.60 $\pm$ 0.01 & 56307.2556 & 2.60 $\pm$ 0.09 & --   \\
--  & 56310.1675 & 2.85 $\pm$ 0.03 & 56310.2441 & 2.58 $\pm$ 0.1 & -- \\
--  & 56312.2388 & 2.72 $\pm$ 0.01 & 56312.1718 & 2.68 $\pm$ 0.1 & --\\
-- & 56327.1409 & 2.51 $\pm$ 0.02 & 56327.0731 & 2.05 $\pm$ 0.05 & --\\
-- & 56333.1279 & 2.41 $\pm$ 0.02 & 56333.1147 & 2.12 $\pm$ 0.09 & --\\
-- & 56335.0700 & 2.51 $\pm$ 0.01 & 56335.0795 & 2.30 $\pm$ 0.03 & --\\
-- & 56340.0047 & 2.61 $\pm$ 0.01 & 56340.1722 & 2.18 $\pm$ 0.1 & --  \\
-- & 56363.0045 & 2.57 $\pm$ 0.01 & 56363.1066 & 2.34 $\pm$ 0.04 & --\\
--  & 54473 & 2.58 $\pm$ 0.03 & 54473 & 2.50 $\pm$ 0.12 & \cite{2008refId0}\\
--  & 54558 & 2.51 $\pm$ 0.08 & 54558 & 2.22 $\pm$ 0.16 & --\\
--  & 54559 & 2.33 $\pm$ 0.06 & 54559 & 2.13$\pm$ 0.10 & --\\
{Mkn\,501}  & 56855.04 & 1.81 $\pm$ 0.02 & 56855.91 & 2.27 $\pm$ 0.10 & \cite{2014-5012020}\\
--  & 56856.86 & 1.74 $\pm$ 0.01 & 56856.91 & 2.01 $\pm$ 0.05 & --\\
--  & 56858.93 & 1.69 $\pm$ 0.01 & 56858.98 & 2.04 $\pm$ 0.04 & --\\
--  & 56861.99 & 1.77 $\pm$ 0.02 & 56861.01 & 2.06 $\pm$ 0.08 & --\\
--  & 56865.04 & 1.73 $\pm$ 0.01 & 56865.00 & 1.97 $\pm$ 0.06 & --\\
-- & 56866.92 & 1.69 $\pm$ 0.01 & 56866.00 & 1.97 $\pm$ 0.02 & --\\
--  & 56867.99 & 1.60 $\pm$ 0.01 & 56867.00 & 1.99 $\pm$ 0.05 & --\\
--  & 56868.92 & 1.74 $\pm$ 0.01 & 56868.01 & 1.93 $\pm$ 0.03 & --\\
--  & 56869.92 & 1.71 $\pm$ 0.01 & 56869.92 & 1.96 $\pm$ 0.07 & --\\
--  & 56032.174 & 1.796 $\pm$ 0.030 & 56032 & 1.85 $\pm$ 0.025 & \cite{2012-501Ahnen_2018}\\
--  & 56036.043 & 1.727 $\pm$ 0.031 & 56036 & 1.88 $\pm$ 0.08 & --\\
--  & 56040.123 & 1.836 $\pm$ 0.033 & 56040 & 2.31 $\pm$ 0.18 & --\\
-- & 56073.333 & 1.671 $\pm$ 0.027 & 56073 & 2.20 $\pm$ 0.07 & --\\
--  & 56074.059 & 1.704 $\pm$ 0.035 & 56074 & 2.01 $\pm$ 0.09 & --\\
--  & 56076.021 & 1.709 $\pm$ 0.028 & 56076 & 2.20 $\pm$ 0.15 & --\\
--  & 56093.988 & 1.796 $\pm$ 0.049 & 56093 & 2.15 $\pm$ 0.12 & --\\
-- & 56095.050 & 1.679 $\pm$ 0.028 & 56095 & 2.05 $\pm$ 0.10 & --\\
\hline
\end{tabular}%
\label{tab:table}%
\end{table*}%

\section{Broadband SED Modelling}
\label{sec:4}
In order to understand the VHE spectral hardening of Mkn\,421, we selected the epoch MJD 56335 during which simultaneous 
observations were available in Optical/UV, X-ray, MeV-GeV $\gamma$-ray and VHE $\gamma$-ray. This epoch was selected since 
the VHE spectrum showed significant hardening. The data corresponding to these energy bands are acquired from \emph{Swift}-UVOT/XRT and \emph{Fermi}-LAT telescopes, while the VHE spectral information  is obtained from \citet{2013Balokovi__2016}.
Due to low photon statistics at $\gamma$-ray energy, the \emph{Fermi}-LAT data is integrated over a period of four days 
from MJD 56334 to 56337.  The integrated $\gamma$-ray spectrum during this period is fitted by log-parabola function 
\begin{equation}
\frac{dN}{dE}=K_0\left(\frac{E}{E_p}\right)^{-\alpha-\beta\log\left(\frac{E}{E_p}\right)}, 
\end{equation}
where $K_0$ is  the normalization, $\alpha$ is the spectral index at the photon energy $E_p$ and $\beta$ decides the curvature at SED peak. 
Our fitting resulted in $\alpha = 1.18\pm 0.44$ and $\beta =0.16\pm 0.15$ for the choice of $E_p$ = 1286.47 MeV. 
During the spectral fit, we assume the spectral parameters of sources other than Mkn\,421 in the ROI do not change
and hence, we froze the model parameters corresponding to these sources to the values given in the 4FGL catalog.
To generate the broadband SED, we obtained the $\gamma$-ray fluxes at five energy bins within 0.1--300 GeV.
The X-ray fluxes are obtained by fitting the source spectrum with absorbed power law model and following the standard 
data analysis procedure as discussed in section \S \ref{Sec:2.1.2}. In case of \emph{Swift}-UVOT, magnitudes were extracted 
from images using tool UVOTSOURCE with source and background regions of radii 5 and 10 arcsec respectively. The observed 
fluxes were de-reddened for Galactic extinctions. The final broadband SED is shown in Figure~\ref{fig:4}. 

\subsection{Emission Model}
The broadband SED is modelled considering synchrotron and SSC emission mechanism from a relativistic non-thermal electron distribution. The emission region is assumed to be a spherical region of size $R$, moving down the blazar jet at relativistic speed. Due to the narrow inclination of the jet from the line of sight of the observer, the emission is significantly Doppler boosted and we denote this Doppler factor as $\delta_D$. The emission region is permeated with a tangled magnetic field $B$ and the relativistic electron distribution loses its energy predominantly through synchrotron and SSC loss processes. Besides these emission processes, there can be also be substantial hadron content in the jet which may produce additional emission, particularly at very high energies. Hence, to model the broadband
SED and the hard VHE spectrum of Mkn\,421, we consider different scenarios of particle distribution involving only leptons and also in combination with the hadrons.

\subsubsection{Broken Power-law Electron Distribution}
\begin{figure}
\includegraphics[width=0.35\textwidth, angle = 270]
    {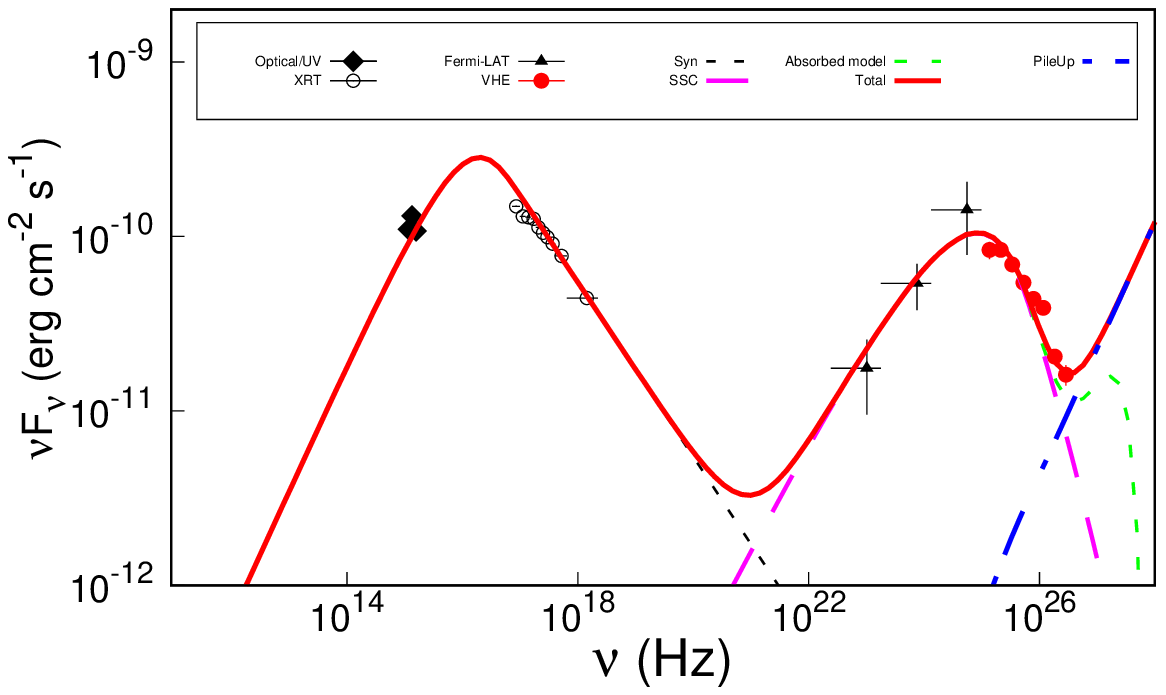}
\vspace{-0.8cm}
\caption{Broadband SED of Mkn\,421 obtained from the observations 
by \emph{Swift}-UVOT/XRT, \emph{Fermi}-LAT and \emph{MAGIC} . The optical/UV,  X-ray, $\gamma$-ray and VHE points 
are denoted by filled squares, filled circles, filled triangles and small filled circles.
The solid red line corresponds to the total model emission emission  involving the contribution from synchrotron, SSC  and Maxwellian pile-up emission. The dotted black, magenta and blue lines correspond to synchrotron, SSC and Pileup emission componets respectively. The dotted green line represents the EBL absorbed model.}
\label{fig:4}
\end{figure}

We assume the non-thermal electron distribution responsible for the broadband emission to be a broken power-law, defined by

\begin{align}
N(\gamma) = 
\begin{cases}
K\gamma^{-p}d\gamma \quad \quad \gamma_{min} < \gamma < \gamma_b\\
K\gamma^{q-p}\gamma^{-q}d\gamma \quad \gamma_b < \gamma < \gamma_{max}
\end{cases}
\end{align}
Here K is the particle normalization with units $cm^{-3}$, $\gamma$ is the Lorentz factor of the electron distribution, $p$ and $q$ are 
the low and high energy power-law indices of electron distribution with $\gamma_b$ the break energy, and  $\gamma_{min}$ and
$\gamma_{max}$ representing the minimum and maximum available energy of the electron distribution. The synchrotron and SSC emissivity functions due to broken power-law electron distribution are solved numerically \citep{Shah_2017,Sahayanathan_2018}. Instead of broken power-law electron distribution, the observed SED can also be explained by the emission from a power-law electron distribution with an
exponential cut-off \citep{2015A&A...580A.100S}. However, in this work, we considered a broken power-law electron distribution which can plausibly originate from multiple acceleration process \citep{2008MNRAS.388L..49S}. The resultant flux at the observer is corrected for cosmological and relativistic effects \citep{1980Natur.287..307B}. The observed SED is reproduced using the synchrotron and SSC model flux with a proper choice of parameters. The source parameters chosen are $B = 0.27 G$, $R =4.5\times 10^{15} cm$, $\delta_D = 26.9$, 
particle energy density $U_e = 1.1\times10^{-2} erg/cm^3$, $p = 1.55$, $q=4.0$, $\theta = 2\, degrees$ and $\gamma_b = 2.6 \times 10^4$. 
To extend the spectrum from radio and VHE band, we  chose   $\gamma_{min} = 10$ and $\gamma_{max} = 10^8$. In Fig. \ref{fig:4}, we show the spectra due to synchrotron and SSC along with the observed SED. The model is able to reproduce the optical/UV, X-ray, and low energy $\gamma$-ray flux; however, at VHE it fails to account for the observed flux (dashed line).
 
\subsubsection{Broken Power-law Electron Distribution with high energy Pile-up}

The hard VHE spectrum of Mkn\,421 cannot be explained by the synchrotron and SSC emission from a broken power-law 
electron distribution. To resolve
this, we introduce in the broken power-law electron distribution an additional quasi steady high energy pile-up at its
maximum available electron energy ($\gamma_{\rm max}$). Such an excess may be possible in the region of particle acceleration when the energy loss rate dominates the acceleration rate for electrons with Lorentz factor $\ge\gamma_{\max}mc^2$ \citep{1998A&A...333..452K,2002ApJ...578..763S}.
We approximate this high energy pile-up as a $\delta$-function given by 
\begin{equation}
\label{eq:eq8}
N_{*}{\left(\gamma\right)} = N_0\; \bf\delta{\left(\gamma - \gamma_{max}\right)}
\end{equation}
and deduce the inverse Compton emissivity due to up-scattering of the synchrotron photons as (Appendix~\ref{sec.app})
 \begin{equation}
 \label{eq:eq10}
 j_{\rm com*}(\nu^\prime) = \frac{N_0c\sigma_T }{4\pi} \;U_{\rm syn}\left(\frac{3\nu^\prime}{4\gamma_{\rm max}^2}\right)
 \end{equation}
where, $U_{\rm syn}$ is the synchrotron photon energy density.

With this additional Compton spectral component and by setting $N_0=10^{-15}\, cm^{-3}$, we were able to reproduce the 
hard VHE spectrum. However, the intrinsic VHE SED will be monotonically increasing until it reaches the 
Klein-Nishina scattering limit. Nevertheless, the observed VHE spectrum do not reflect this feature since it is significantly
suppressed by the absorption due to EBL interaction. We also attempted to fit the broadband SED using different values of $\gamma_{max}$. However, we observed that lowering the value of $\gamma_{max}$ causes the pileup emission component to shift to lower energies, resulting in a corresponding shift of the peak intensity towards lower energy regions. Consequently, this caused the model to deviate from the observed hard VHE data points. To ensure consistency with the observed data, we fixed $\gamma_{max}$ at $10^8$ while reproducing the broadband SED points.
The modified SED along with the observed VHE spectrum is shown in Figure~\ref{fig:4}.

The high energy pile-up in the particle spectrum will also cause excess in synchrotron emission. The synchrotron emissivity 
due to this pile up will be (in the rest frame of th emission region)
\begin{align}
	j_{\rm syn*}^\prime =\frac{\sqrt{3}e^3 B N_0}{16  mc^2} f\left(\frac{\nu}{\nu_c}\right)
\end{align}
where,
\begin{align}
	f(x) &= x\int\limits_x^\infty K_{5/3}(\xi)\, d\xi  \nonumber \\
	&\approx 1.8 \, x^{1/3}\, e^{-x}
\end{align}
and 
\begin{align}
	\nu_c=\frac{3\gamma_{\rm max}^2 e B}{16 m c}
\end{align}
For the chosen set of parameters, the observed synchrotron spectrum will peak at 
\begin{align} 
	\nu_{\rm  syn *}&\approx\left(\frac{\delta_D}{1+z}\right)\gamma_{\rm max}^2 \left(\frac{eB}{2\pi mc}\right)\\
	&= 1.98 \times 10^{23} \quad {\rm Hz}
\end{align}
and the flux in $\nu F_\nu$ representation at this frequency is found to be $\approx 1.13 \times 10^{-15}$ erg/cm$^{2}$/s.
This flux is sub-dominant to the contribution from the broken power-law electron distribution and hence will not 
deviate the low energy spectrum.

    \begin{figure}
\includegraphics[width=0.35\textwidth, angle = 270]
    {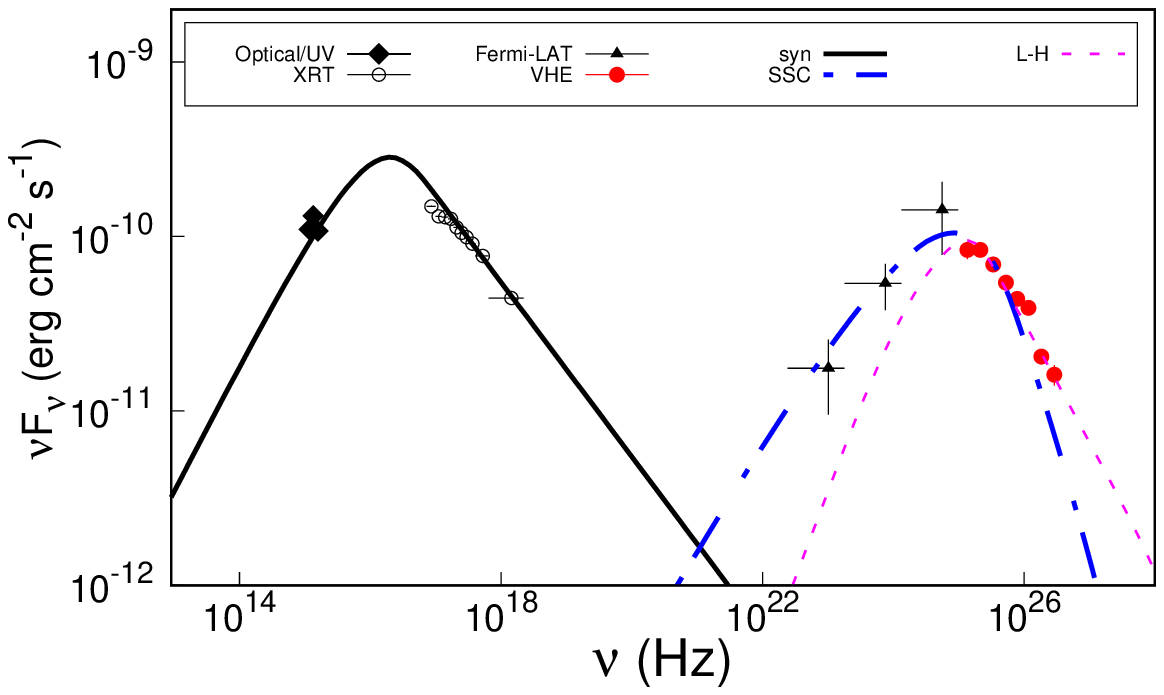}
\caption{Broadband SED of Mkn\,421. The points are same as defined in Figure \ref{fig:4}.  The black solid, and blue  dashed line correspond to total  model emission involving the contribution from synchrotron and SSC emission. The dashed pink dotted line corresponds to the emission  involving the contribution from leptohadronic process.}
\label{fig:leptohadronic}
\end{figure} 

\subsection{Lepto-hadronic model}

The possible alternative considered here to explain the VHE excess is the contribution of energetic protons whose contributions remain sub-dominant in lower energies. To model this scenario, we consider the photomeson process where the relativistic protons interact with the synchrotron photons present in the relativistically moving emission region. Here we considered the production of neutral pions ($\pi^0$) in photomeson process and the decay of $\pi^0$ into two $\gamma$-photons. The production of charged pions ($\pi^\pm$) which subsequently produce electrons and positrons, is not considered here. 

To estimate the $\gamma$-ray spectrum produced in the photomeson process through the decay of neutral pions we used the formalism developed by \citet{Kelner_2008}. The energy distribution of $\gamma$-rays produced due to the decaying 
$\pi^0$-meson is given by \citep{Kelner_2008}
\begin{equation}
\frac{dN_{\gamma}}{dE_{\gamma}} = \int f_p(E_p) n_{ph}(\epsilon) \Phi_{\gamma}(\eta, x) \frac{dE_p}{E_p} \, d\epsilon 
\end{equation}
where $f_p(E_p)dE_p$ and  $n_{ph}(\epsilon)d\epsilon$ are the number of proton and number of synchrotron photons densities in the energy ranges $dE_p$ and $d\epsilon$ respectively. $\Phi_{\gamma}(\eta, x)$ is the angle averaged cross-section of the p$\gamma$-interaction. The dimensionless quantities $\eta$ and $x$ are given by \citep{Kelner_2008}
 \begin{equation}
	 \eta = \frac{4\epsilon E_p}{m_p^2c^4}, \qquad x = \frac{E_\gamma}{E_p}
 \end{equation}

The details of the parametric form of the function $\Phi_{\gamma}(\eta, x)$ is given in \citet{Kelner_2008}. The estimated broadband SED  using lepto-hadronic model along with the spectral data of Mrk\,421 is shown in figure \ref{fig:leptohadronic}. 
As can be seen that it can also explain the VHE component of the the source. In earlier studies, the hard TeV spectrum has also been explained using hadronic and leptohadronic models. \citet{2022MNRAS.512.1557A} have explained the hard TeV spectra of six extreme  
BL Lacs using two zone leptohadronic model. The VHE $\gamma$-ray emission is well explained by the photo-hadronic processes in the spherical zone near the BL Lac core while as low energy spectra(X-rays) is due to synchrotron process in the far zone. The 
accelerated protons in the near zone interact with the photons from the sub-relativistic pair plasma. The 2018 TeV detection from  EHBL 2WHSP J073326.7+515354 by MAGIC collaboration was better explained by hadronic model\citep{2019MNRAS.490.2284M}. In 
such models $\gamma$-ray emission is due to protons in the jet or by the emission from particles formed during proton-photon interaction
\citep{1993A&A...269...67M,Aharonian_2000,bou2013EPJWC..6105003B}

\section{Summary}
\label{sec:5}

We investigated the variation of intrinsic VHE spectral indices with X-ray indices of two nearby blazars viz. Mkn\,421 and Mkn\,501. 
The simultaneous X-ray and  VHE  observations of Mkn\,501 show that the VHE spectrum is steeper than the X-ray spectrum. 
This may indicate that the SSC process, responsible for the $\gamma$-ray emission, fall in the Klein-Nishina scattering regime. If 
the scattering process happens at Thomson regime, one would expect the VHE index to be equal to the X-ray spectral index as the same electron population
is responsible for the emission at these photon energies. The scatter plot of the simultaneous X-ray and VHE observations of Mkn\,421, 
on the contrary, show a hard VHE spectra and this cannot be attributed purely due to SSC process happening neither under Thomson nor Klein-Nishina scattering regime. 
This hints that at the VHE energy of Mkn\,421, additional emission process may be active. It is important to note that the effect of EBL  can be significant at multi-TeV energies. If the VHE index strongly depends on the flux at multi-TeV, then it indicates that the observed VHE spectrum is steeper. Consequently, the effect of EBL would  harden the intrinsic  VHE spectrum further than the ones estimated in this work . As a result, our conclusion that the VHE spectrum in Mkn\,421 is harder than the X-ray spectrum remains valid.

To account for the additional emission process at VHE, we explore two plausible scenarios in this work. First, we consider a quasi-steady high energy 
pile-up of electrons at the maximum energy of the broken power-law electron distribution \citep{2021MNRAS.508.4038H}. Under Fermi acceleration, efficient radiative loss at high energy can lead to such quasi-steady pile-ups in the accelerated electron distribution \citep{2002ApJ...578..763S}.
Approximating this pile-up of electrons as $\delta$-function, we reproduce the hard VHE spectra of Mkn\,421 by considering the Compton up-scattering of 
the low energy synchrotron photons. In the second case, we consider the photo-meson process as the additional source of VHE emission. 
Along with the electrons, protons can also be accelerated to ultra-relativistic energies that can interact with the low energy synchrotron photons
and produce pions. The emissivity due to the decay of this neutral pion into two $\gamma$-ray photons is used to account for the excess VHE emission.
We find the observed hard VHE spectrum in Mkn\,421 can be well explained under both these scenarios. Possible detection of neutrino emission from the source 
have the potential to scrutinize  these models.

One can avoid the hard VHE spectrum of Mkn\,421 by claiming that the EBL absorption is overestimated. Theoretically  this is possible by incorporating the  Lorentz invariance violation scenario \citep{1999ApJ...518L..21K}, which  likely takes place  above 2 TeV \citep{2001APh....16...97S}. However,   the hard spectra problem in distant blazars occurs  at sub-TeV energies. Other possibility to resolve the hard VHE spectra is by considering the VHE photons as secondary $\gamma$-rays produced  by the interactions of cosmic ray protons with background photons \citep{essay2011ApJ...731...51E}. Alternatively, one can also argue that the $\gamma$-ray photons oscillates as light axion-like particle close  to the source and finally  converts back to photons before reaching the Earth \citep{essay2011ApJ...731...51E}.  However,  these  scenario would 
be considerable if the source is located at a large distance and also demand the existence of exotic particles or very low magnetic fields  ($\sim 10^{-15}$G).
Therefore,  in a more realistic astrophysical scenarios, the formation of the non-thermal particle distribution is more likely reason for the hard VHE spectra. 

 Besides the leptonic models, a number of alternate explanations to the hard $\gamma$-ray spectrum has also been explored in the literature. For instance \citet{lepCerruti_2015} has explained the hard TeV spectra using  lepto-hadronic modelling similar to the one considered in this work. 
They modelled the low energy component of blazar SED as the synchrotron emission from a relativistic electron distribution; while, the 
high energy component is the result of SSC and pion-induced cascade process. Particle acceleration due to 
magnetic re-connection has also been considered to explain the hard spectra. Such acceleration process is capable of producing a hard particle
distribution with index $\sim $ 1.5 \citep {mag2014ApJ...783L..21S}. In spite of various explanations for the hard VHE spectrum from sources,
the observational evidence to converge to the right model is currently unavailable. Probably, a detailed temporal modeling of the blazar
light curves at different energy band along with VHE have the capability to eliminate the inconsistent models. With the advent of high 
sensitivity instruments, operating at VHE like Cherenkov Telescope Array (CTA), it is expected to provide a better understanding 
regarding the $\gamma$-ray emission from blazars.

\section{Acknowledgements}
 AM, SS, NI and ZM are thankful to the Department of Atomic energy (DAE),Board of Research in Nuclear Sciences (BRNS), Govt of India via Sanction Ref No.: 58/14/21/2019-BRNS for the financial support. ZS is supported by the Department of Science and Technology, Govt. of India, under the INSPIRE Faculty grant(DST/INSPIRE/04/2020/002319).

\section{DATA AVAILABILITY}
The multi-wavelength data used in this work is publicly available at the data browse section of the respective observatories. Codes used in this work will be shared on request to the corresponding authors.




\bibliographystyle{mnras}
\bibliography{refrences} 




\appendix 
\section{Inverse Compton emissivity due to mono energetic Electron Distribution}
\label{sec.app}
The inverse Compton emissivity in the rest frame of the emission region can be approximated as \citep{coopi1990MNRAS.245..453C, 1998ApJ...509..608T}
\begin{equation}
\label{eq:A2}
J(\nu^\prime) = \frac{c\sigma_T}{8 \pi } \int_{\nu_1}^{\nu_2}N{\left[\left(\frac{3\nu^\prime}{4\nu_0^\prime}\right)^\frac{1}{2}\right]}{\left(\frac{3\nu^\prime}{4\nu_0^\prime}\right)}^\frac{1}{2}{\frac{U_{\rm ph}(\nu_0^\prime)}{\nu_0^\prime}}d\nu_0^\prime
\end{equation}
Here, $N$ is the underlying electron distribution and $U_{\rm ph}$ is the target photon density which is being up scattered. The primed 
quantities are measured in the rest frame of the emission region. 
If we assume the electron distribution as a $\delta$-function
\begin{equation}
\label{eq:A4}
N{\left(\gamma\right)}
= N_0\; \bf\delta{(\gamma - \gamma_{\rm max})},
 \end{equation}
In order to obtain the inverse Compton emissivity $J_{*}$ due to scattering of the synchrotron photons, we 
recast the $\delta$-function over the $\nu_0^\prime$. The final emissivity can be obtained from 
equation (\ref{eq:A2}) as
 \begin{equation}
 J_{mp}(\nu^\prime) = \frac{N_0c\sigma_T }{4\pi} U_{\rm ph}\left(\frac{3\nu^\prime}{4\gamma_{max}^2}\right)
 \end{equation}
 The observed flux on earth can be obtained from this emissivity after accounting for the relativistic boosting and 
 cosmological effects \citep{1980Natur.287..307B,1995ApJ...446L..63D}.


\bsp	
\label{lastpage}
\end{document}